# Latent Viral Marketing, Concepts and Control Methods

Alon Sela
Department of Industrial Engineering
Tel Aviv University
alonsela@post.tau.ac.il

Dmitri Goldenberg
Department of Industrial Engineering,
Tel Aviv University
dmitrig@mail.tau.ac.il

Irad Ben-Gal
Department of Industrial Engineering,
Tel Aviv University
bengal@tau.ac.il

Erez Shmueli
Department of Industrial Engineering,
Tel Aviv University
shmueli@tau.ac.il

*Abstract* – **Numerus works that study the spread of information in social networks include a spreading mechanism in which a set of nodes is initially infected (i.e. seeded), followed by a viral process, which spontaneously spread the message through the nodes of the network. These models are used to describe the spread of rumors as well as the spread of new products and services. In reality however, it is quite rare that a product or service spreads through a social networks solely by viral forces. It is more common, that the spreader would invests a continuous active effort of its sales representatives in order to enhance the spread. The Latent Viral Marketing Model is a spreading model that fits this reality. Along the description of the model, the paper continues and proposes a simple Scheduling of Seeding heuristics, which recommends the node to seed at each period. A large study of empirical simulations shows that under a wide range of realistic initial conditions, the Scheduling Seeding method improves a product adoption rate by 29%-94% in comparison to existing state-of-the-art seeding methods.**

## I. INTRODUCTION

Social networks provide a powerful communication platform that is known to significantly influence social and historical events. For example, its impact can be observed in President Obama`s first elections [1], [2], in the Arab Spring uprising [3], in the spread of propaganda by ISIS recruiters [4], or in the US election of 2017. The importance of social networks is mainly a result of their efficiency in quickly spreading information among many users. These traits are the reasons why political parties, terror organizations and commercial firms, use social networks to a growing degree.

Generally, it is believed that information spread follow the Linear Threshold Model [5]. According to this model, first, the spreader selects several chosen nodes and seeds them (where the act of a seeding [6], reflects an intentional infection of nodes). Then, a viral process begins, where the information spreads through the nodes of the social network and users infect each other's. Such an act of infection, can be performed for example, if a user writes on his Facebook wall a new message, which is later seen by the user`s friends. Then, the user`s friends can chose to send this message to their own friends. In each step. A user can alternatively send the message directly to a single friend or send it to a group of friend. In Twitter, the user spreads a message by Retweeting a received message, or simply by twitting a new tweet that contains a relevant link. The follower of the user would then have this message presented in their tweets time line, and can open the link. If they find the message interesting, they can retweet it, thus the message will appear on the time line of their followers.

The spreading methods mentioned above are all active methods. This implies that users must perform an action (such as a retweet or post on a wall) in order to spread the message to another. In each step, users invest an effort (i.e. "work") toward spreading the message. Such a method of information sharing is desirable to many commercial firms. These firms seek to harness the viral forces of these many "free spreaders" and have these spreaders invest their effort to spread the commercial firm`s products or services.

The firms wish to gain an almost unlimited source of free workers that spread their product or services. Unfortunately, very few products spread solely by viral forces, and most firms still need to employ sales and marketing departments to promote their products and services. The low ability spread products and services by viral mechanisms is not due to the low importance of social forces in the act of purchasing. In fact, one`s social connections are known for many years to have an immense influence on one`s personal decisions making. The first social psychologists, Asch [7], Milgram [8], Granovetter [9] and Zimbardo [10], revealed the importance of social influence as a key factor influencing one`s attitudes and values. Social proximity in a social network predict tendencies that were believed to be genetic. For example, the tendency for obesity, smoking, or even the tendency of being happy [11]. If such internal traits spread through the links of the social network, shouldn`t we expect that recommendation for a new products would spread as well?

In practice, encouraging customers to invest efforts to spread a commercial product is not always as easy as it seems [12]. First, it is required that these customers would like the product a lot! This is a critical requirement in any effort for successful viral marketing. This is true, since an influential customer, which actually dislikes the product, will influence the spread in a negatively [13]. Furthermore, customers do not usually like to promote commercial firms by their own good will. A few relatively recent works [14], [15], [16], [17], and [18] have shown that the tendency that a customer will spread a commercial products is lower than previously believed [19], [20], [21]. These works observed the lengths of information cascades in large data sets and found them rather short and shallow. It seems as the vast majority of messages never spread through thousands of users, but rather through a relatively small number of users. Since large information cascades are rare, it is also rare that a product or service spread solely by viral forces. An external aid to spread is usually required for most products or services.

The following work proposes a model of information spread, which considers social influence in a more realistic way. The method recognizes the importance of social forces, but does not expect to gain "free workers" from it. An expectation to have a customer actively spreading a commercial product or services is not too realistic. The customers influence on his friends is important, but the influence is not sufficient and is not active, it is hidden, and latent.

As a motivating example, let us consider a setting in which a given company aims to promote the sales of one of its products. The company's sales representatives might contact customers and offer them to purchase the product by phone, or by using an equivalent advertising platform. If a customer purchases the product, he / she might tell some of his / her friends (network neighbors) about this purchase. We assume that these friends will not actively contact the company to purchase the product by themselves, but would rather keep the positive recommendation latent in their minds. However, if contacted by the company's sales representatives within a certain period, the positive recommendations accumulated on the product, as provided to them by their friends, is likely to influence their likelihood to purchase the product. If the sales representative address the customer long after the customer`s friends` recommendations have been heard, the customer is less likely to purchase. The company thus needs to decide which users to approach and at what points in time in order to utilize its sales budget in an efficient manner, while taking into account the latent influence; e.g. the effect of the user`s friends recommendations as is accumulated in the customer`s minds.

This work fits the scenarios above. We define the *Latent Viral Marketing Model* (LVM), and a related seeding method, the *Scheduling Seeding Heuristics* (SSH), which increases the number of successful seeding attempts in the above scenario. The work adds a stochastic aspect to the deterministic Scheduled Seeding method that were developed and presented in our previous works [22], [23], and adds the realistic latency to the sequential seeding process described in [24]. Thus, the work follows a setting that we believe fits real life scenarios to a greater degree.

According to a large set of simulations, SSH significantly improves the number of successful seeding attempts in scenarios similar to the LVM model described above, in comparison to existing state-of-the-art seeding benchmark approaches.

These benchmark approaches mainly focus on careful selection of nodes with high centrality measure [25], such as PageRank, Eigenvector Centrality, or simply the node's Degree Rank in the initial seeding process. More specifically, the SSH reaches an average improvement rate of 23%-153% in the number successful seeding attempts (depending on initial conditions), and in some extreme cases reaches an improvement of up to 10 folds.

The next section includes a brief background on information diffusion models through social networks and in particularly, on the Linear Threshold model. The background section is followed by an in-depth description of the proposed LVM information spread model followed by the SSH seed selection heuristics. We present the results of various simulations experiments and summarize the study by a concluding paragraph.

## II. MOTIVATION

### A. The Linear Threshold Model

One of the most popular models in the field of viral marketing is the Linear Threshold model [5]. This model starts when an initial set of nodes is first infected, followed by a viral process model which simplifies social influence. According to the Linear Threshold model, the viral spread will flow if $\sum_{w \in W_v} b_{v,w} \geq \theta_v$, where $W_v$ denotes the set of infected neighbors of $v$, and $b_{v,w}$ denote the weights; i.e. the social influence that $w$ activates on $v$. If the total influence reaches a threshold $\theta_v$, node $v$ changes its state and becomes infected.

Plotting the total number of infected nodes versus the elapsed time, while applying the Linear Threshold viral process, creates a plot that often resembles a sigmoid function. The number of infected nodes slowly increases at the beginning of the process, then after enough infected nodes accumulate; it increases sharply, up to the point where most nodes are infected. Then, when additional uninfected node becomes scarce, the speed of infection slows down, and the slope decreases.

Similar sigmoid spreading curves represent many physical phenomena of spreads, such as for example a forest fire or virus epidemic. In a forest fire, after an intentional ignite (i.e. seeding) the fire spread is first slow. Then, as the fire grows, it quickly spreads by its own forces to the rest of the forest. In this period, in many cases, the fire can burn large parts of the forest in a short period. At the end, when much of the forest is already burn, the fire slowly decays since unburnt trees are infrequent, until it completely vanishes.

This sigmoid growth function, while fitting numerus natural spreading phenomena, does not seem to fit the spread of ideas through social networks.

### B. Growth of Actual Information Cascades

A growing body of works, [15], [17], [26], [27], [28], which analyzed several large social networks data sets, claim that large information cascades are rater rare. Most information cascades only spread through two people, even fewer spread to three. For example, a spread of a message to five friends occurs in only 1/8 of the messages, and a spread of a message 3 times in a vertical cascade (i.e. an initial message that is spread to a friend, and this friend spreads it again to his/her friends), only occurs in 1/16 of the messages. An even larger spread, for example a vertical spread of 8 steps, was found to only occur in between 0.01% - 0.001% of the messages. These studies were replicated over different social networks; and capture an important aspect of information cascades. While people collect information consistently, they do not always actively diffuse it further to their friends.

Information cascades differ from the spread of biological virus in their selective nature. While collecting the information might be similar to receiving a virus, and people do collect much of the information they receive, information spread is more selective. People tend to distribute information cautiously and not repeat everything that they have heard to everyone.

This is one of the reasons, why it is rather rare that a company succeeds to distribute its products with no additional effort, simply by using a viral process. In contrast to a virally based strategy, most companies need to spend a lot of effort (and budget) to actively help the spread of their products. Most companies need to construct brands names through commercial communication methods, they employ sales personnel, and they actively promote social network marketing strategy. The conventional Linear Threshold model does not address the every day's scenario, in which a company invests substantial effort to promote a product or a service. This is why the LVM model is required.

*C.    Assumptions Underlying the Latent Viral Marketing Model*

The first works on information spreading through social networks compared the spread phenomena to the spread of viruses. The SIR model is the basic model of virus spread. It has been studied for over a hundred years, with an excellent work summarizing the vast studies in the field found in [29]. Unlike the spread of a biological virus, social norms influence the adoption of ideas. The rate of acceptance of a certain idea in one's social circle predicts the likelihood for adopting it. Social norms are indeed integrated in the Linear Threshold model, which defines the probability of infection as the sum of intentioned neighbors' weights. Although the theoretical importance of this work, along similar [30], [31] information cascades works is evident, these works does not fit a case of commercial products or services in which the spread requires continues effort of marketing and sales departments.

In order to fit the Linear Threshold model to these scenarios, we must first change the deterministic nature of the model. Another required modification is the clear separation between the seeding stage and the viral stage, which does not fit the reality of a commercial product spread. The investment of an entire budget in a single and initial period is in many cases impossible. Most commercial firms have limited call centers capacities, and can only reach a limited number of customers per day. Lastly, many spreading models assume that if a certain number of neighbors of a person adopt a product, the person will adopt it as well. While this might be true, in many cases one might be willing to adopt a product or service following good recommendations from friends, but he does not adopt it simply because he is too busy to actively reach the company and acquire the product or service. Nevertheless, if reached by a sales person, he is likely to adopt the product or service.

The Latent Viral Model provides a new framework, which we hope overpass the obstacles mentioned above. It assumes from one hand that nodes accumulate social information, and that this accumulated information is a major factor in the adoption decision. However, in contrast to the previous Independent Cascade philosophy, it assumes that new nodes cannot become infected solely by a viral process. Instead, an external effort of a sales representative is required before a node actually becomes infected. Thus, the question of seeds allocation is relevant not only in the initial stage, but also along the entire spread process.

The work follows previous works [22], [23], [32] on the topic, and studies the stochastic aspect of the Scheduling Seeding problem. It includes a deeper inspection of the ideas presented in these previous studies.

Correspondingly, the challenge is to decide on what node it is worthy to invest the seed and at what period. This decision is based on the feedbacks received from previous seed attempts along the current social network structure. Such feedbacks include the knowledge of customers that have already adopted the product, along those who have not. As shown in the result section, when using the LVM model, the success rates of the seeds trials grow if the Scheduling Seeding Heuristic (SSH) is used. A more formal description of the model followed by the heuristics used to select the seeds is presented in the next section.

III.    THE PROPOSED MODEL

*A.    The Setting*

Let us consider a company with good visibility on the social network of its clients. The company wants to offer its customers a new service or product (we use the term service or product interchangeably), and offers it through its sales representatives. The company seeks to achieve the highest possible number of customers that adopt the new service, and allocates a limited budget, denoted by $B$, to promote this goal.

If the company offers the service to a customer; let us say $v$, the customer might accept or reject the offer with a certain probability $p$. This probability is affected by the adoption rate of the service by the customer`s social circle, as further explained below in eq.(2). In a case where the customer refuses the offer, subsequent offers in the near future will only annoy the customer, and therefore the product would not be offered again to the same customer. In such a case the customer`s state is considered to be in a "Seeding Failed" state. The social influence is such that if a customer accepts the offer, the customer is likely to influence his / her friends for the next $t^{inf}$ periods. In this case, after $t^{inf}$ periods have ended, he the customer changes from a state of being infected and infectious, to a state of being infected but not infectious. This state change reflects the retention loss, or the loss of interest in the message due to information overload [33]. The possible states of a customer are denote by $St_v$ and as defined below.

(1) $$St_v = \begin{cases} 0 - & Non-Infected \\ 1 - & Infeted\ and\ Infectious \\ 2 - & Infected, Non-Infectious \\ 3 - & Seeding\ Failed \end{cases}$$

The various states changes $St_v$ that customer $v$ might follow, are presented in the states transition scheme in Fig. 1.

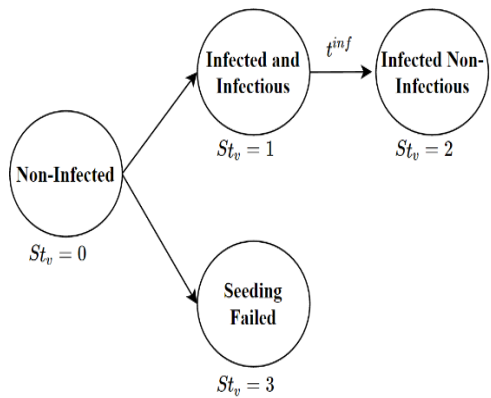

Fig. 1. Possible States and the causes for states changes of a customer.

B. *Defining the Probability of a Successful Seed Attempt*

The probability that customer $v$ accepts an offer is affected by the social pressure executed on the customer, as well as the attractiveness of the proposed product or service itself. We therefore define a maximal probability for an infection (adoption) by the proposed service or product, and denote it by $p_{Max_v}$. This parameter depends on the type of product or service, and can usually be estimated from past data. For example, the probability of accepting an offer for three months free cable TV service without any commitment might be rather high, while the probability of accepting an online purchase of a new luxury car is low. The probability of accepting the proposed offer follows eq.(2), where $N_v^+$ is the set of indexes of the infected ($St_v = 1$) neighbors of customer $v$, and $\theta_v$ is the minimal number of infected neighbors at $p_{Max_v}$. This formulation fits the results appearing in Asch`s conformity experiments [7], [34], which inspected the probability of conforming to norms as a factor of group size (see Fig.2).

(2) $$p_v = p_{Max_v} \cdot \left(\frac{min(\theta_v,\ |N_v^+|)}{\theta_v}\right)$$

In his works, Asch inspected how group size influences the probability of conforming to the opinion of the majority. As the coalition of this majority grows, the conforming probability grows almost linearly, until a certain coalition size (see Fig.2 right), where we denote this value by $\theta_v$. Note that larger coalitions above $\theta_v$ will not increase the likelihood of conforming any more. This influence function is plot in Fig.2. (left) which presents the probability of conforming as it was copied from Ash`s original article on social conformity [34]. In comparison, the same right image of the figure presents the approximated function as defined in eq.(2). In both figures, the x-axis represents the number of people adopting the opinion, and the y-axis indicates the equivalent probability of acceptance, with the peak y-axis equal to $p_{Max_v}$.

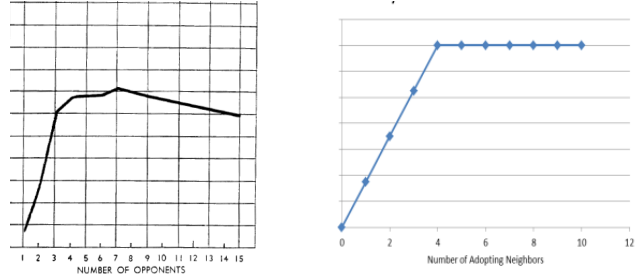

Fig.2. Social Influence Function based on Asch's conformity experiment. Right, figure directly copied from Asch`s article **[34]**, as compared to left figure which presents an approximation function as defined by eq.(2), with parameters $\boldsymbol{p_{Max_v}} = \boldsymbol{0.35}$ and $\boldsymbol{\theta_v}$=4.

C.   *The Pre-Seeding and Seeding Processes*

According to the definition of the acceptance probability as defined in eq.(2), if there is not even one infected node in the entire social network graph $\boldsymbol{G} = (\boldsymbol{V}, \boldsymbol{E})$, the term $|\boldsymbol{N_v^+}| = 0$; $\forall v \in \boldsymbol{V}$. It follows that $p_v = 0$; $\forall v \in \boldsymbol{V}$, and if course, in such a case, no seed trial would succeed. To prevent of being trapped in such a zero attractor, prior to the spread process, we define an initial set of infected nodes and set them to state $\boldsymbol{St_v} = 1$; i.e. infected and infectious. These nodes are chosen randomly from $\boldsymbol{V}$, and this pre-seeding infected set is defined by $\boldsymbol{F^{init}}$. The relative size of $|\boldsymbol{F^{init}}|$ is usually small, and consists of less than 1% of the nodes. Furthermore, the infection times of the nodes in $\boldsymbol{F^{init}}$ are set such that each of these nodes has a different initial infection time, thus they do not change from $St_v = 1 \to St_v = 2$ at once but rather gradually.

Following the initial setting of $\boldsymbol{F^{init}}$, the seeding process starts. The process includes $B$ seeds attempts, which are performed on selected nodes. Assuming each seeding attempts costs exactly one unit of budget, and $M_s$ nodes can only be seed at each period, these limitation fits real scenarios in which call centers can only make a limited number of phone calls per day due to their work hours constraints.

The Scheduling seeding algorithm nodes selection include three steps. First, the algorithm chooses a set of nodes that are not yet infected, but have at least one infected neighbor. These are the potential candidates for the seeds. Second, it computes an "attractiveness" score for each of these candidates. Third, in each period, $M_s$ nodes with the highest scores are seeded.

After the seeding is performed, the simulative stage "decides" if the seeds are accepted or rejected. The seeding trial succeeds or fails with a probability $p$, which as defined according to eq.(2). After the seeding of each period, relevant parameters and state changes are executed for the relevant nodes in the network. These include the calculation of $N_v^+$ for each node, as well as changes of states for odes that require such a change. This process ends when the entire budget is depleted, or when all the nodes in the network becomes infected. Once the process ends, the ratio of successful seeding is computed, simply as the number of seeding successes per seeding trials.

D. *The Seeding Scheduling Heuristics (SSH) Score Computation.*

The Seeding Scheduling heuristics recommended the seed trials by computing their LVM score at each step. The score is based on the expected value of the node being seeded, and reflects the probability of an occurrence of an event $p(\psi)$, multiplied by the utility $U(\psi)$ of the event. The event $\psi$, is defined as the success in a seed of node $v$.

The utility function gained from $\psi$ is constructed from two separate parts. First, the success seeding of $v$ has a utility of one additional infection node. Second, to this term, an additional term is added as the utility gained from the increased probabilities of future successful seeding of the uninfected neighbors of $v$. Since $v$ is now infected, its neighbors are now easier to seed. The first term, the utility gained from the successful seeding of $v$ simply equal to 1. The second term; the increased probabilities of uninfected neighbors of $v$ is defined as the sum of changes over all nodes $u \in N_v^-$ where $N_v^-$ denote the non-infected neighbors of $v$. Assuming the event $\psi$ occurs, this second term is the value of the nodes $N_v^-:\psi$ minus their current value $N_v^-:\bar{\psi}$. Thus, the utility from seeding trial to $v$ is $U(v) = \sum_{u \in N_v^-} p(\psi) \cdot U(\psi) - \sum_{u \in N_v^-} p(\bar{\psi}) \cdot U(\bar{\psi})$, where $\psi$ is the states of the neighboring nodes $u \in N_v^-$ after the seed of $v$ succeeded, and $\bar{\psi}$ is their states before the seed of $v$ succeeded. This is a recursive formulation, since $U(\bar{\psi})$ is actually unknown.

The computation method to calculate this score, is performed recursively, and is defined for a depth of 3 recursion levels in eq.(3). The recursive computation of the score, for a depth of $k$ iterations, is presented in the following algorithm, and is later named as the seeding strategies "picky_social_<k>"; i.e. "picky_social_0", "picky_social_1" and "picky_social_2", where *picky_social_0* computes the score simply as the term $p(v)$, picky_social_1 as $p(v) * \{1 + \sum_{u \in N^-(v)}[p(u)]\}$ and picky_social_2 as the full in eq.(3).

$$(3) \quad Score(v) = p(v) * \{1 + \sum_{u \in N^-(v)}[p(u) * (1 + \sum_{w \in N^-(u)} p(w))]\}$$

This attractiveness score is computed recursively as defined in the pseudo-code below

**The SSH Scoring Algorithm**

*Function Social Score(v,G,k):*
*# input: v - relevant node, G - Graph, k - Levels (Social_0, Social_1, Social_2)*

$Set\ p(v) = p_{Max_v}(\frac{min(\theta_v,\ |N_v^+|)}{\theta_v})$   # Probability of infection of $v$ in current time step
$if\ Levels = 0:$
    $return\ p(v)$  # Level 0 - Greedy score
$else:$
    $set\ Score = 1$
$for\ u\ in\ N^-(v)$               # go over all v's non – infected neighbors
    $score = score + Social(u, G, Levels - 1)$
$return\ p(v) * Score$

To clarify the above recursive method, note that the expected value of $v$ itself is $p(v)*1$, and the expected value of $u$ (where $v$ is the 1st circle) which weren't infected is $p(v)*p(u/v)*1$, meaning an occurrence of both events (successful infection of $v$ and afterward successful infection of $u$ which is based on a new probability calculation. Similarly, the expected value of $w$ (where $v$ is in the 2nd circle, and $u$ is in the 1st circle, is : $p(v)*p(u/v)*p(w/u,v)*1$), which is the formulation defined in eq.(3).

While the above function computes the scores at any level of $k$, as further seen in the result section, there is a tradeoff between the effort to foresee and the time of computation. In most cases, it seems as the right balance is in one single level of depth, that is, in setting the parameter *k=1*. At this depth of recursive, the results are rather good, but the additional computation complexity dramatically increases. In the next section, we present the methods used to evaluate the efficiency of the above SSH scorings under the LVM modeling, followed by the results from these sets of experiments.

## IV. EVALUATION

### A. Experimental setup

We set an empirical experiment in order to compare the performance of suggested and existing benchmark seeding heuristic. The experiment was executed on a Linux Cluster with 128 GB RAM memory, where the simulations were run on a single Intel 2.7GHz CPU. Each simulation instance started with a setup of the initial condition, which included a selection of a pre-simulation infected set $F^{init}$ as defined above. The infected time of this set were generated from a uniform distribution, such that there would not be a sharp decline in the infectious nodes at period $t = t^{inf} + 1$.

As the simulation started, in each period, a seed was offered to a single node, whereas the selection of the seeded node was based on different heuristic rules. Each seeded node could accept or reject the seed with a probability based on its surrounding nodes according to eq.(2), and the node`s state function changes were calculate at each discreet period. The simulation instance ended when the entire budget was used, then, the final seed success to failure rate was calculated.

The results examine the different seeding strategies across changing dimensions of initial parameters, such that at each set of simulations, a single dimension was examined across a wide range of values. The other parameters were set to their default values, which were in most cases the median of the range. During each simulation run, the SSH seeding recommendations under the LVM simulations was compared to the benchmark seeding methods, throughout the entire parameter space, while running each parameter combination for at least 400 replications. The entire parameter space as used in the simulation experiment is presented in Table 1 below.

**Table 1 – Simulation Parameter Space[1]**

| Parameter | Values |
|---|---|
| Network (see Table 2 for further details). | **Sampled Citation network**, Slashdot Network, Sampled EuEmail network, WikiVote Network, Epinions Network, Enron Network |
| Network size (sample # nodes) | 5000, 10000, 50000, **100000**, 500000, 1000000 |
| Initially infected population size | 50, 100, **200**, 500, 1000 |
| Max Budget | 50, 100, **200**, 500, 1000 |
| Threshold | 3, 4, **5**, 6, 7 |
| Maximal Probability | 0.1, 0.3, **0.5**, 0.7, 0.9 |
| Infection Time (Time of Oblivion) | 10, 20, **50**, 100, 200 |
| Seeding method (our method and the benchmark methods) | Random, GEC, Picky Random, Picky_GEC, LVM (Picky_Social_0, Picky_0ocial_1, Picky_0ocial_2) |

Three different SSH seeding recommendations under the LVM simulations scheme i.e. Picky_Social_0, Picky_social_1, Picky_social_2 were compared to four benchmark methods which included (1) Random, (2) GEC, (3) Picky Random, (4) Picky GEC as further defined in the next section. These simulations were executed on different networks as defined in Table 2 below.

---

[1] Default value of each dimesion is marked with bold. The value is used when other parameters change.

Table 2 – Networks[2] Used in Simulation

| Network | Number of Nodes | Avg Degree | Avg Clustering | Network Type |
|---|---|---|---|---|
| Citations | 1000000 | 2.83481 | 0.039113922 | Sampled & Undirected |
| Citations | 500000 | 4.057372 | 0.060063242 | Sampled & Undirected |
| Citations | 100000 | 7.60482 | 0.136068811 | Sampled & Undirected |
| Citations | 50000 | 8.19712 | 0.160465584 | Sampled & Undirected |
| Citations | 10000 | 6.809 | 0.200986075 | Sampled & Undirected |
| Enron | 36692 | 10.020222 | 0.49698256 | Full & Undirected |
| Wiki_vote | 7115 | 28.323823 | 0.140897846 | Full & Undirected |
| Slashdot | 82168 | 14.179072 | 0.06034486 | Full & Undirected |
| Euemail | 100000 | 1.56686 | 0.034104364 | Sampled & Undirected |
| Epinions | 75879 | 10.694395 | 0.137756373 | Full & Undirected |

*B.     The Benchmark Methods*

The evaluation of the SSH seed recommendations method for the LVM scheme, was compared to the four benchmark methods below.

(1) *Random* – Randomly choosing one uninfected node to seed at each time step.
(2) *GEC* – Choosing the one uninfected node with the highest Eigenvalue Centrality measure at each time step.
*(3) Picky Random* – Choosing a random uninfected node from the nodes which have at least one infected and infectious neighbor.
(4) *Picky GEC* – Choosing an uninfected node with the highest Eigenvector centrality from the nodes that have at least one infected and infectious neighbor.

These benchmark methods were compared to these three SSH social heuristics.

(1) *Picky_Social_0* - Choosing a non-infected node with the highest value of $p(v)$ at each time step, see SSH Scoring Algorithm with k=0.
(2) *Picky_Social_1* - Choosing the non-infected node having the highest value of Score(v) as defined in the first part of eq. (3), and the SSH Scoring Algorithm with k=1.
(3) *Picky_Social_2* – Choosing the node having the highest value of Score(v) as defined in eq. (3) and the SSH Scoring Algorithm with k=2.

In the simulations, we first assume that the values of the parameters $\theta_v$ and $p_{Max_v}$ are known. In the second set of experiments, we assumes that we only know the mean and variance of these parameters, along their distribution. The means are denoted by $\mu_\theta$ and $\mu_{p_{Max_v}}$, and the variance are denoted by $\sigma_\theta$ and $\sigma_{p_{Max_v}}$ accordingly. The real value of these parameters for each node were generated prior to each simulation run, and were not preliminarily known to the SSH recommendation algorithm.

V.     RESULTS

*A.     Centrality of the nodes chosen to be seeded*

The Eigenvector Centrality measure of a node (as well as its PageRank score) are considered as a good proxy for a node`s ability to spread information. The main concept behind the LVM Scheduling scheme is that it is not only the centrality of the node that defines its importance, but rather its tendency to accept the information at any precise period of time.

---

[2] Networks were downloaded from SNAP, http://snap.stanford.edu/data/index.html then sampled.

Before presenting the entire sets of results, we thus first examine the nodes chosen for the seeding at each period in regards of their Eigenvector centrality. This inspection allows us to validate that the success of the SSH method is not simply because it prefers seeding central nodes.

We compare the Eigenvector Centrality of each seeded node along time, when using the GEC method, a method that allocates the seed to the relevant non-infected nodes by their Eigenvector Centrality scores, to the scheduling methods by the LVM method (i.e. named Social).

A comparison presenting the centrality of the selected nodes in the LVM method is presented in Fig. 3Fig. **3**. below. The SSH Social method, (blue line) allocates seeds to nodes with relatively lower average Eigenvalue Centrality, as compared to the GEC method. While in a single run (presented in the interior plot of the figure), we can see that nodes with high Eigenvalue Centrality can be seeded in a rather later stages, the average Eigenvalue Centrality (presented in the exterior plot) of the nodes along time is substantially lower for the SSH.

It can be concluded from those first results that the LVM method does not allocate the seeds to central nodes, but rather that it allocates them to nodes that are of high importance at any current point in time. As is further seen, the SSH selection of nodes does not only allocate seeds to less central nodes (which in reality might be easier to reach), but rather results in final success rate that is substantially higher that of the benchmark methods for any given budget.

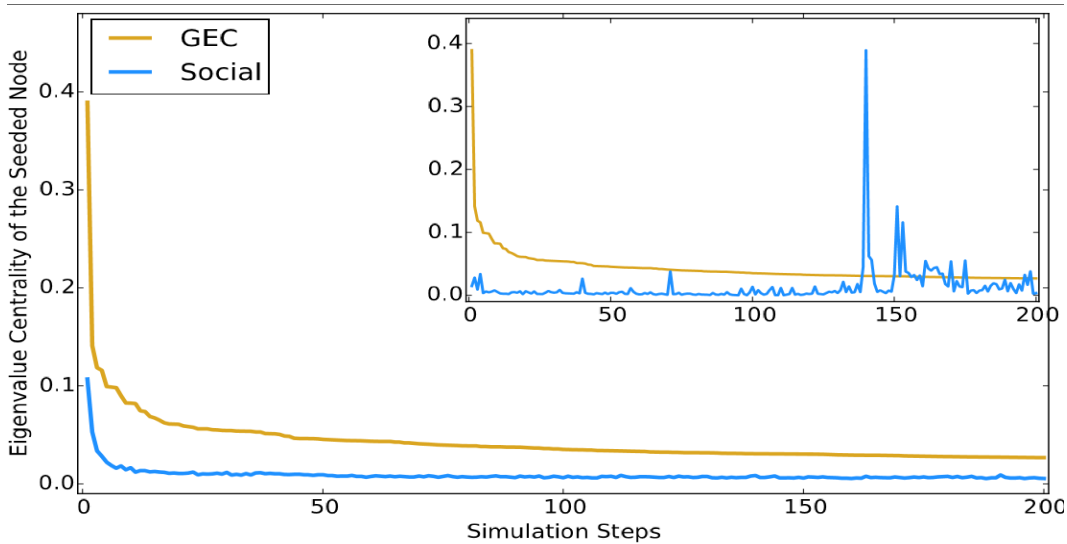

Fig. 3. Eigenvector Centrality of the nodes chosen for seeding. Allocation of seeds to influential nodes; i.e. highest Eigenvector Centrality (orange), as compared to the allocation of seeds by the SSH social methods (blue). The x-axis is the time of seed attempt, while the y-axis is the Eigenvector Centrality of the node on which the seeding attempt is performed.

## B. *Comparing the LVM with the Benchmark methods*

We start by comparing the SSH method to the benchmark methods, for different network sizes. As can be seen in Fig. 4 below, the social methods (blue bars) outperform the benchmarks methods by almost twice. For all the different seeding methods, the Social 2 method seem to reach the best results, followed by the social 1 and the Social 0 methods. In comparison, the benchmark method of Picky GEC, a method that allocates seeds to nodes with the highest Eigenvector Centrality in condition that these nodes already have at least one infected neighbor, only succeeds at about ~13% as compared to ~20% success rates for the Social method. Note that the GEC and the Random methods practically used by many commercial firms that do not include the network structure of their clients in their marketing efforts. The success rates in these methods are far lower.

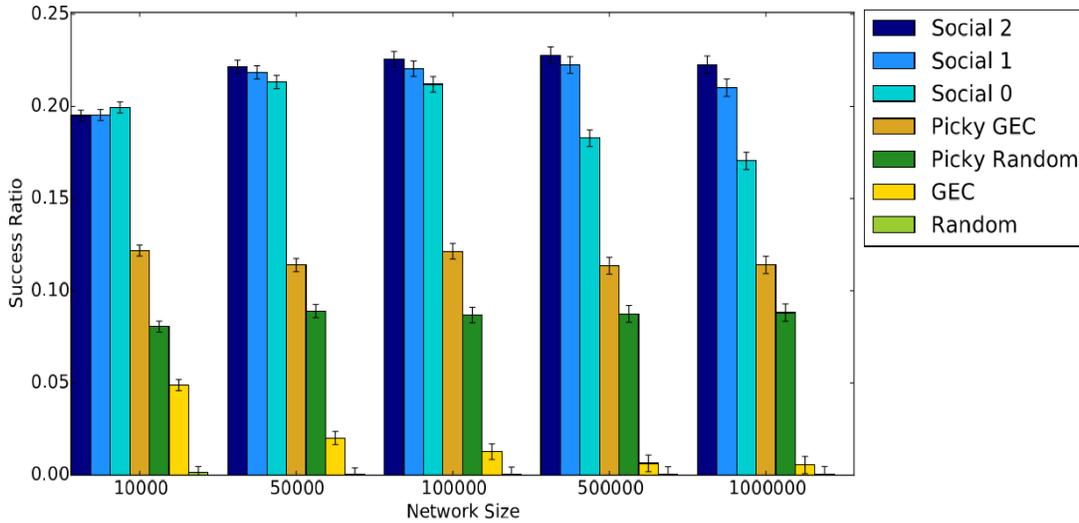

Fig. 4. Comparison of SSH Scheduling method (blue scale) to the benchmark methods for different sizes of networks.

The results in Fig. 4 are on sampled citation networks of different sizes. We follow these results and validate them for different networks, on diverse average degrees and Clustering Coefficients.

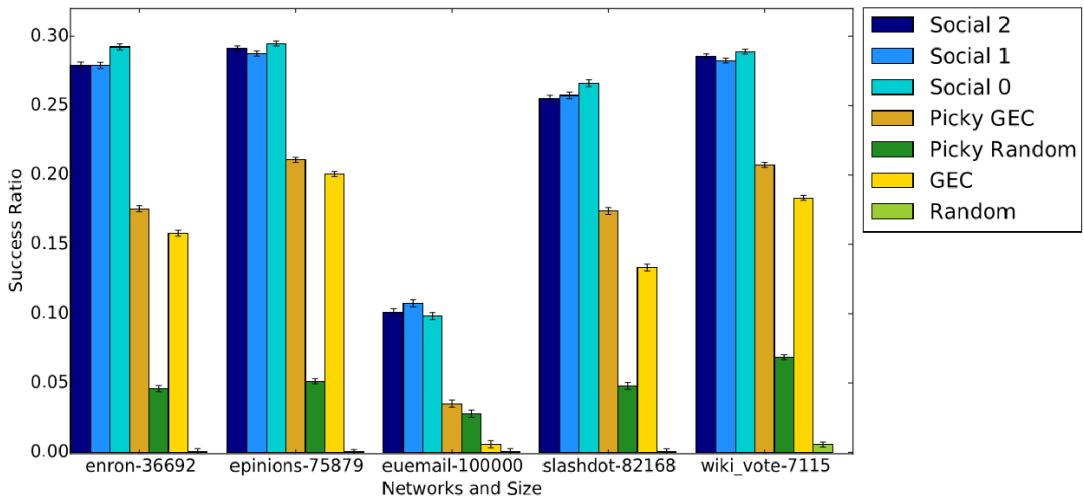

Fig. 5. Comparison of SSH Scheduling method (blue scale) to the benchmark methods for different networks topologies.

As seen in Fig. 5, the results are mainly similar. Note that the euemail-100000 network has substantially lower success rates as compared to the other networks. To understand these results, we need to look at the average degree of this network and compare it to the average degree of the other networks (see Table 2). While euemail-100000 network has an average degree of 1.56, the other networks have an average degree of 10.7 or higher. The low degree in the euemail-100000 network reduces the probability of any seed success, since in the LVM model, the number of infected neighbors is a major factor influencing the probability of a successful seed, when the network is sparse, and this probability is accordingly low.

Note also that in these results, unlike the case of the generated network, the Social 2 is not always the best method. Similarly, Social 1 is not always better than Social 0. It seems as in reality, when the network topologies differ, in many cases it is better to use the simple Social 0 and Social 1 heuristics over the more complex Social 2 heuristics which tries to plan forward for two steps in advance.

The GEC methods seed nodes with high Eigenvector Centrality in earlier stages. This might create a larger influence at early stages and improve the acceptance rates later on. In order to inspect the temporal aspect of the spread, we measured the success rates of the different seeding methods along the time axis. These results, as presented in Fig. 6, indicate a growth in the success rate along the time axis. The growth is larger in the Social seeding methods (blue), as compared to the non-social methods (yellow or green). The growth in the success ratio seems to follow a log like function, since the y-axis is the ratio and not the absolute number. These results imply that for growing budgets (growth in time) we expect a constant benefit from using the social methods as compared to the benchmark methods.

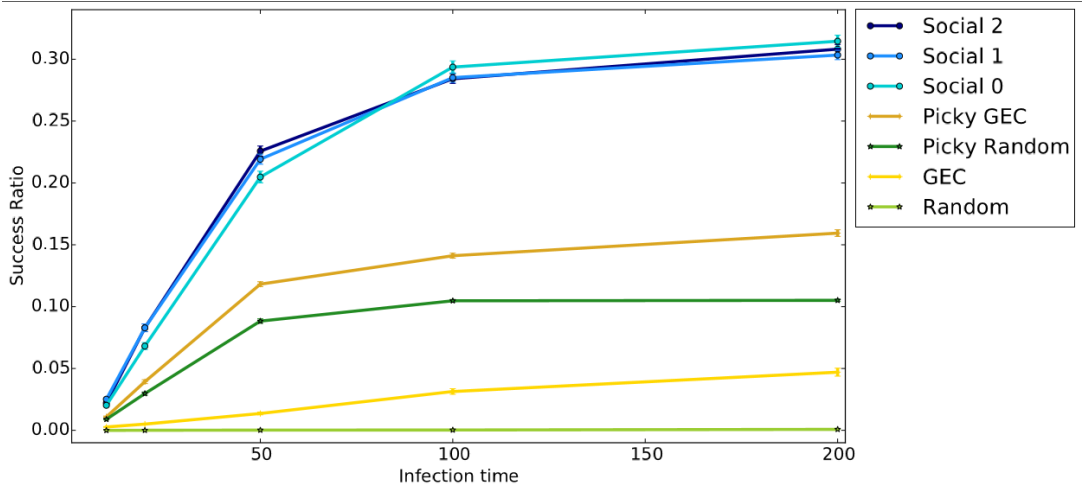

Fig. 6. Temporal comparison of the aspect of the SSH Scheduling method (blue scale) to the benchmark methods

As described in the proposed model section, prior to the seeding attempts, the states of nodes in $F^{init}$ were set to $St_v = 1$. We inspect the influence of the size of $F^{init}$ on the different seeding methods.

As demonstrated in Fig. 7. below, a larger initial population in $F^{init}$ (x-axis) improves the relative utility of the Social methods. When the initial set $F^{init}$ consists of only 50 infected nodes, the Social methods succeeds in the seeding 16.6%-18.4% of the seeding attempts. In comparison, the Picky GEC methods succeeded in the seeding of 11.5% and the Picky Random only succeeds in 9% of the cases. This is an improvement of 44% for the Social methods. As opposed to this initial setting of $F^{init}$, if $F^{init} = 1000$, the social methods succeeds in seeding 29%-28.1% of the seeding attempts, while the Picky GEC and Picky Random only succeeded in 14.5% and 9.1% which is an improvement of 94%. Thus, the improvement of the Social methods over the next best methods grow from 44% to 94% as $F^{init}$ grows.

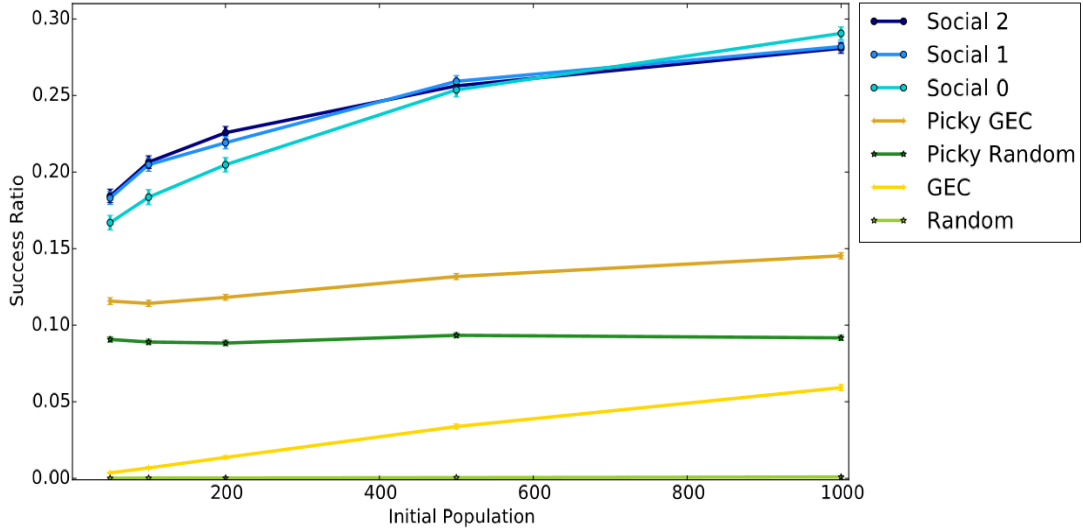

Fig. 7. Influence of number of infected nodes prior the seeding on the different Scheduling method. Social marked by blue scale, as compared to GEC and Random methods marked Yellow and Green accordingly.

## C. Initial Parameter Estimations

In the LVM model, the probability of a successful seeding, as defined in eq.(2) is based on the three parameters. The first is the number of adopters in the social circle of a node $v$; denoted $|N_v^+|$. This parameter is known, assuming that we know exactly which of the node`s neighbors have adopted the service or product that is being spread. The two other parameters are $p_{max}$ and $\theta_v$ which are unknown. We will first present an analysis, which inspect the influence of these parameters on the results, followed by an inspection of the more realistic scenario, in which the values of these parameters are not known, but they are estimated through their statistical properties.

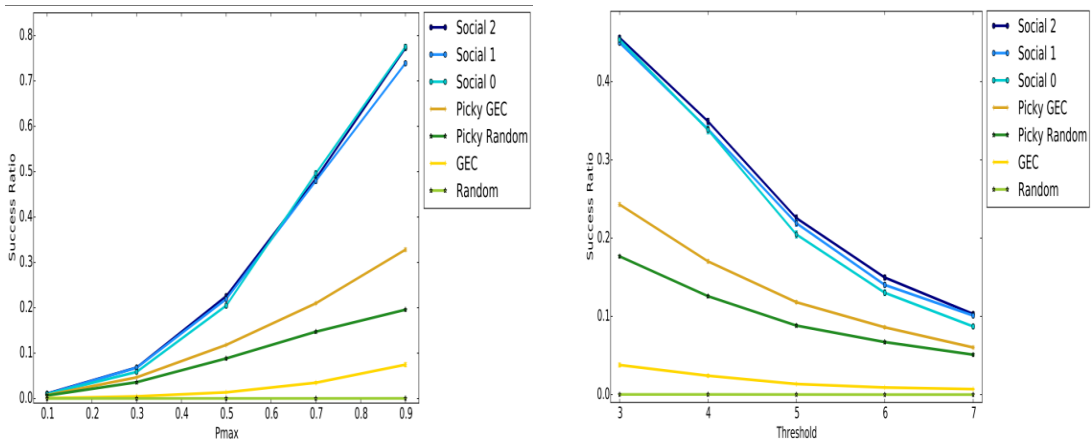

Fig. 8. Influence of $p_{max}$ (left figure) and $\theta_v$ - the threshold (right figure) on the success rates of the Social methods (marked blue lines) compared to the benchmark methods (marked yellow and green lines).

The influence of $p_{max}$ and $\theta_v$ on the results can be observed in Fig. 8. It is clear that higher values of $p_{max}$ (left image), only improves the efficiency of the Social methods (marked by blue lines) as compared to the other benchmark methods (marked by green or yellow lines). This result make sense. A product or service that have a large value of $p_{max}$ are those that have a larger probability of purchase if one`s friends have purchased. For example, such products can be trendy products for teenagers or kids, where the social influence plays a large role in the desirability of the product. For these products, it would be reasonable to assume that the LVM method, a strategy that better incorporates the social aspect of the purchasing decision would be beneficial over more static approaches, which only include the network topology.

Regarding the threshold value $\theta_v$ as presented in the right figure, higher values of $\theta_v$ represent products where one need to accumulate more adopting neighbors before one reaches a purchasing maturity. Products or services that fit the category and are expected to have high values of $\theta_v$, are products or services where one tends to accumulate much information prior the purchasing maturity. These can be important (and costly) decisions such as buying a new car or new home. In these important decisions, where one tends to invest one`s time and effort in profound inquiries prior the purchasing decision, the social aspect is less dominant. While the trend seem to continue such that the social LVM methods are always preferred, these are decisions where the success ratio is also small. Note that such cases as expensive decisions, the social methods (when one consults as many as 7 friends) is 8%-10% for the Social methods, as compared to 6% for the picky GEC method. This represent an improvement of at least 33% for the Social methods over the best of the other methods, which in a case of an expensive product or service is a very good result.

*D. Inspecting the Simulations Space with Unknown Parameters*

The results described in the section above assumed that the values of $p_{max}$ and $\theta_v$ are known. Of course, this is not true. At the best, the distribution of these parameters can be estimated, but the individual parameter for each node is never known. For this reason, we conducted another set of experiments and inspected the performance of our scheduling method under the LVM for unknown values of $p_{Max_v}$ and $\theta_v$. In these experiments, the means and standard deviations of $p_{Max_v}$ and $\theta_v$ were known, but the true value of these parameters for each node was not revealed to the Scheduling algorithm.

We thus first generated value for $p_{Max_v}$ and for $\theta_v$ prior the run, then run the different seeding methods while not letting the algorithm know the values of the parameters for each node. In each run, the Scheduling algorithm simply generate a possible value for $p_{Max_v}$ and $\theta_v$ from their means, standard deviations and distributions and continued to search for the best node as if their values were known. We assumed the parameters distribution is a Normal distribution, and inspected the influence of growing rates of errors rates with growing standard deviations of these parameters.

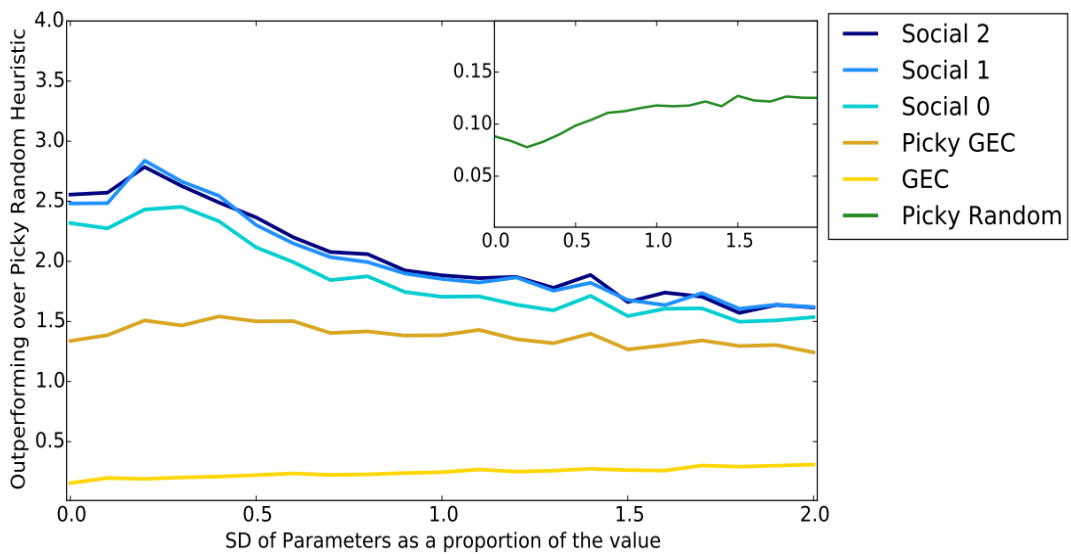

Fig. 9. Social methods improvement rates for different degree of uncertainly

As seen in Fig. 9, the growing degree of uncertainty (x-axis) of the real values of $p_{Max_v}$ and $\theta_v$, results in a decreasing performance of the social LVM methods, as compared to the random method. We set the Picky Random method (which randomly selects nodes that have at least one infected neighbor) as a comparison line, and only inspect the degree in which each distinct method performs better than the random method. Note that for the random heuristics (inner plot); an addition of noise actually improves the performance of the method. If the values of $p_{Max_v}$ and $\theta_v$ have larger variance, it implies that in some cases $p_{Max_v}$ and $\theta_v$ would be low. In these cases, if the nodes selection is random, the probability of a seed success is high. Since we compare the performance of each heuristics to this random heuristics, which grows with the addition of more noise, we expect that more "noise" to result in a less accurate plan of the Social method as compared to the random method. It can be seen, that even for the high levels of a standard deviation of as much as 2 standard deviations, the worst Social method (i.e. Social 0) still performs better than the random by 153%, and the best benchmark method (i.e. Picky GEC) only performs better than the random by 124%. This represents an improvement of at least 23% for the Social method over the best other benchmark method. Furthermore, for smaller levels of standard deviations, (cases where we can better estimate the parameter values) the improvement of the Social methods as compared the other methods is substantially higher.

### E. Aditional Unknown Parameters of Minimal Probability of Adoption

The previous section inspected the behavior of the model when the parameters $p_{Max_v}$ and $\theta_v$ were unknown. These parameters represent the uncertainty related to the highest possible probability of seed success, in a case where there are many infected neighbors. There is nevertheless, another source of uncertainty, which was not addressed in the LVM model. This is the case of a product adoption when none of one`s friend have never adopted it. It is clear that while social influence is an important aspect in the purchasing decision, there are cases where one purchase a product or service that none of one`s friend have purchased.

As seen in eq.(2), when none of one`s friends have adopted the product or service, the value of $|N_v^+| = 0$, and the probability of adoption is accordingly 0. This difficulty in the model, of course needs a correction. In order to correct it, we redefined eq.(3), as below, by adding a minimal value $p_{Min_v}$ to the term.

(4) $$p_v = p_{Min_v} + (1 - p_{Min_v}) \cdot [p_{Max_v} \cdot (\frac{\min(\theta_v, |N_v^+|)}{\theta_v})]$$

The term $p_{Min_v}$ thus represent the a priori probability of a node accepting a seed, when none of its neighboring nodes have accepted it.

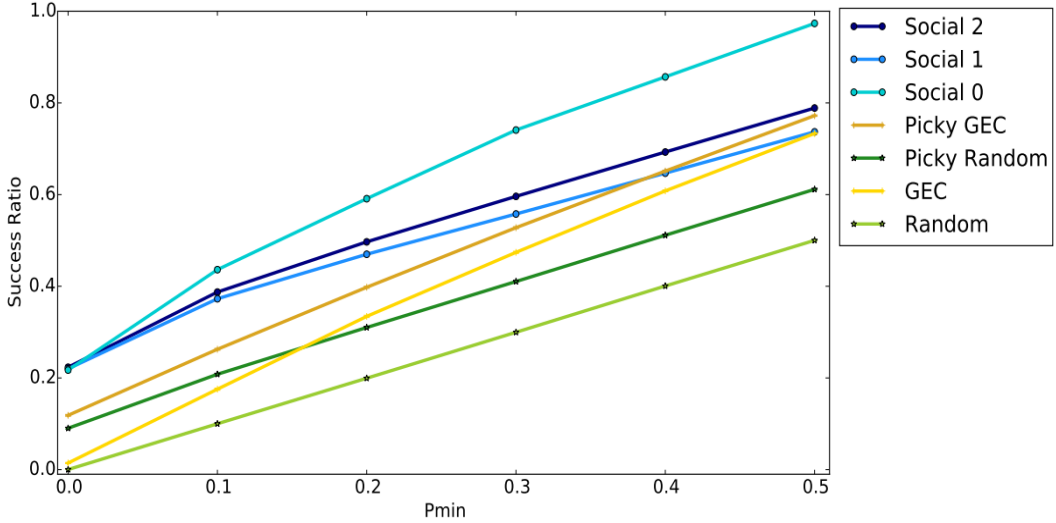

Fig. 10. Influence of $p_{Min_v}$ on the success rates of the Social methods (marked blue lines) as compared to the benchmark methods.

The additional term $p_{Min_v}$ to the LVM model, reveals two interesting properties of the LVM model and the Social heuristics. First, note that when $p_{Min_v}$ is added to the LVM model, the Social 0 method outperform the other Social methods. This trend can be explained by the low ability of the more complex Social Algorithm to correctly predict the seeds success when noise is added. Furthermore, note that when the value of $p_{Min_v}$ is $p_{Min_v} \leq 0.4$ the Social methods are still better than the other methods. In contrast, when the value of $p_{Min_v}$ is $p_{Min_v} > 0.4$ the Picky GEC methods gains better results over the Social methods.

These results define the region where the Social methods is expected to gain better results, and enable a better decision when to use the Social methods and when to use the GEC methods. With this in mind, it is important to note that the Picky GEC method does not simply allocate seeds to nodes according to their Eigenvalue Centrality, but rather restricts the nodes allocations to nodes that have at least one infected neighbor. It this includes some type of feedback on what node adopted the offer. If this feedback is ignored, then the correct comparison needs to be the GEC regime and not in the Picky GEC regime. In this method, seeds are allocated to nodes according to their Eigenvector Centrality without concerning their neighbors' state at all. In this case, only when 50% of the purchasing decision is personal ($p_{Min} > 0.5$) it is better to use the GEC methods over the Social methods.

### F.  Run Time of the Social Methods

The different Social methods represent a growing degrees of future planning effort. While the Social 0 method is fully greedy, the Social 1 tries to plan one step ahead, and the Social 2 plans two steps ahead. Although the SSH scoring algorithm, as previously presented can be used with growing degrees of future planning, we did not find sufficient improvement in more than 2 steps plan ahead. This is important if considering the fact that when the networks size grow, the computational cost of the plan ahead grows accordingly. Furthermore, in many cases, a trial to plan for the far future might result in trial to seed nodes that are influential in the long term, but have lower probability of accepting the seed in the short term. Such a strategy can result in lower final success rates since these influential nodes seeding simply fails.

Note that the computation cost of computing the Eigenvector Centrality measure for very large networks is also rather expensive in computational time. As seen in Fig. 11 below, when the network size grow near 800,000 nodes, the computational cost of the most expensive Social method, i.e. Social 2 is already better than that of computing one single time the Eigenvector Centrality measure. As contrast, in networks of sizes of $|n| > 800,000$ nodes, the computational cost of the Social 1 and the Social 0 is still negligible. As much as a network of size $|n| > 1,000,000$ nodes is still a very small network, the runtime is still less than 1 minute, thus it does not seem as the runtime in the Social methods is a real problem.

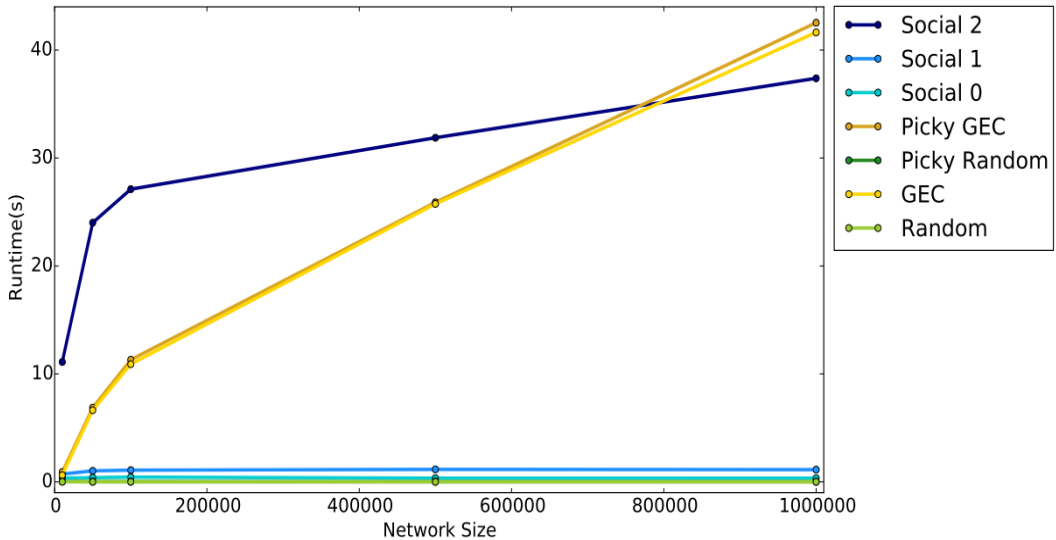

Fig. 11. Run time of different seeding methods for varying network sizes

## VI. CONCLUSION

Many works that study information cascades in social networks, consider these cascades as a phenomenon by which information virally spreads by its own force through the links of the network. Unlike the spread of biological viruses that can be carried passively by agents and infect a significant portion of the network, information cascades are usually much shorter and long cascades are rather rare [17], [15], [14]. These results do not necessarily imply that social forces lost their importance but rather that people information spread is more selective, and does not necessarily fit the use of an SIR model of virus spread.

We propose a new information spread model, in which agents, e.g., sales representative of a company, communicate with network members, e.g., potential clients, and offer them a new product or service. The probability that a client accepts such an offer is based on the acceptance levels of its neighbors.

Since contacting a client includes some financial cost (limiting the number of clients that can be approached at once), the company has to select which members to approach and at what time, in order to increase the total adoption rate in the network.

The proposed Latent Viral Marketing Model and its recommendation method for customer selection, sees influential nodes, as nodes that are most likely to accept an offer at each period and thus influence others.

In a large series of simulated experiments, we show that the proposed method increases the adoption rate in 23%-153% (depending on the initial conditions), over the best-known method, which seeds the nodes by their Eigenvector Centrality measure.

Having said that, it is important to note that the method is applicable to products that have a viral characteristic. These are products or services where a substantial part of the purchasing decision is based on social influence. In products or services where social forces are not important, it might still be better to use the old method of selecting nodes that have high Eigenvalue Centrality measures.

The work contribution can be summarized along three different axes. First, we believe that the LVM spread model better fits real-world scenarios of products adoption, where products spread relies on an effort of a sales department, and seldom spread with no external force added. In these cases, this work directs the sales personals, where and when to contact each possible customer. Second, the proposed model demonstrates the importance and the high potential of a Scheduled Seeding approach, while restricting the scenarios to the cases where this method is expected to be useful, as well as those where it is not. Third, we offer a simple, yet a powerful method (by the SSH algorithm), that can be easily applied in disserve situations of marketing of trendy product, where social forces are of high importance.